\documentclass[aps,groupedaddress,prd,preprint]{revtex4}
\usepackage{graphicx}
\begin{document}
\title{Production and Decay of Sneutrino and Squarks at Lepton-Hadron Colliders }
\author{A.T. ALAN$^{1}$ and S. SULTANSOY$^{2,3}$}
\address{$^{1}$Department of Physics, Faculty of Sciences and Arts, Abant Izzet Baysal University, 14280 G\"{o}lk\"{o}y, Bolu, Turkey}
\address{$^{2}$ Department of Physics, Faculty of Sciences and Arts, Gazi University, 06500 Teknikokullar, Ankara, Turkey }
\address{$^{3}$ Institute of Physics, Academy of Sciences, H.Cavid Ave.33, Baku, Azerbaijan}
\date{\today}
\begin{abstract}
We investigate the potentials of future high energy lepton-proton
colliders to detect supersymmetric particles in the charged
current type $lp \rightarrow \tilde{\nu_{l}} \tilde{q}X$ ,
$l=e,\mu$, reactions. We also study their decays by using the mass
spectrum given in the Technical Design Report of ATLAS
Collaboration (SUGRA Point 6 corresponding to large tan$\beta$).
\end{abstract}
\maketitle
\section{Introduction}
In the context of extensions of the Standard Model (SM), R-parity
conserving Minimal Supersymmetric one (MSSM) is the most studied
scenario of physics beyond the SM [1]. Therefore, searching for
supersymmetric particles is going to be an important part of
experimental programs of future colliders. Meanwhile, because of
the clearest signature, the associated production of the squarks
and sleptons is the most important process in the search for SUSY
in lepton-hadron collisions. In this paper we reanalyzed the
associated squark and sneutrino production in lepton-hadron
collisions [2] for Linac$\otimes$LHC and $\mu$p colliders with the
center of mass energies $\sqrt{S}=5.3$ TeV and 3 TeV, respectively
[3]. We do not consider THERA [4] and LEP$\otimes$ LHC [5] because
of unsufficient center of mass energies $(\sqrt{S}\sim 1 TeV)$.
\section{Calculation and numerical results}
In lepton-proton interactions the R-parity conserving process
$lp\rightarrow \tilde{\nu}_{l} \tilde{q} X$ proceeds via charged
gaugino exchange in the t-channel of the subprocesses
$lq\rightarrow \tilde{\nu}_{l} \tilde{q}$ .The polarized
subprocess cross sections can be written as follows:
\begin{center}
\begin{equation}\label{1}
\hat{\sigma}^{LL}_{RR}=\frac{g_{w}^{4}}{16\pi\hat{s}}\cdot\frac{\tilde{m}_{\chi}^{2}\Delta}{(\tilde{m}_{\chi}^{2}-\tilde{m}_{\nu}^{2})
(\tilde{m}_{\chi}^{2}-\tilde{m}_{q}^{2})+\hat{s}\tilde{m}_{\chi}^{2}}
\end{equation}
\end{center}
\begin{center}
\begin{equation}\label{2}
\hat{\sigma}_{RL}^{LR}=\frac{g_{w}^{4}}{16\pi\hat{s}^{2}}[{(\tilde{m}_{\nu}^{2}+\tilde{m}_{q}^{2}-2\tilde{m}_{\chi}^{2}-\hat{s})}
\ln(\frac{\hat{s}-\tilde{m}_{\nu}^{2}-\tilde{m}_{q}^{2}+2\tilde{m}_{\chi}^{2}-\Delta}{\hat{s}-\tilde{m}_{\nu}^{2}-\tilde{m}_{q}^{2}+2\tilde{m}_{\chi}^{2}+\Delta})-2\Delta]
\end{equation}
\end{center}
where
$\Delta=[(\hat{s}-\tilde{m}_{\nu}^{2}-\tilde{m}_{q}^{2})^2-4\tilde{m}_{\nu}^{2}
\tilde{m}_{q}^{2}]^{1/2}$, $\hat{s}$ is the center of mass energy
of the incoming lepton and quark and $g_{w}$ is the weak coupling
constant.  $\tilde{m}_{\nu},\tilde{m}_q$ and $\tilde{m}_{\chi}$
denote masses of sneutrino, squark and chargino, respectively.

The mass matrix for charginos in the $(\tilde{W},\tilde{H})$ basis
is given as
\begin{center}
M$_{\chi}$=$\left[\begin{tabular}{c c}
  $\tilde{m}_{w}$ & $\sqrt{2}m_{W}\sin{\beta}$ \\
   $\sqrt{2}m_{W}\cos{\beta}$ & $\mu$
\end{tabular}\right]$
\end{center}
 so that gauginos and Higgsinos mix in general to form charginos
 $\tilde{\chi}_{i}$ because of broken SUSY. In SUGRA inspired models
 $\tilde{m}_{w}=\tilde{m}_{g}(\alpha_{w}/\alpha_{s})$ and $\mu$ is
 of order $\tilde{m}_{g}$, therefore, the lighter chargino is going
 dominantly be gaugino. For numerical evaluations we use SUGRA
 Point 6 spectrum from ATLAS TDR [6]. The reason we choose this
 point is the large value of $\tan\beta=45$, which is natural in
 MSSM, because of the relation $\tan\beta=m_{t}/m_{b}$ coming from
 flavor democracy hypothesis in three family case [7]. The mass
 spectrum of SUSY particles is presented in Table I.

By folding subprocess cross-section $\hat{\sigma}$ over the quark
distribution functions f$_{q} (x,Q^{2})$ (in
$l\bar{d}\rightarrow\tilde{\nu}_{l} \bar{\tilde{u}}$ and $\bar{l}
\bar{u}\rightarrow\bar{\tilde{\nu}_{l}} \bar{\tilde{u}}$ cases
sea-quark distributions should be used) [8], we obtain the total
cross-section for the production of a particular sneutrino-squark
pair in lepton-proton collision at the center of mass energy
$\sqrt{S}$;
\begin{center}
\begin{equation}\label{3}
\sigma(lp\rightarrow\tilde{\nu_{l}}\tilde{q}X)=\int_{x_{min}}^{1}dx
\hat{\sigma}(lq\rightarrow\tilde{\nu_{l}}\tilde{q})f_{q}(x,Q^{2})
\end{equation}
\end{center}
where $x_{min}=\frac{(\tilde{m_{\nu}}+\tilde{m_{q}})^2}{S}$. In
figure 1 we plot the total cross-section versus center of mass
energy of lepton-proton collisions using the mass values given in
Table I. Numerical values of corresponding cross-sections and
event numbers for Linac$\otimes$LHC and $\mu$p colliders are
presented in Table II.
\begin{figure}[h]
  \centering
  \includegraphics[width=12cm]{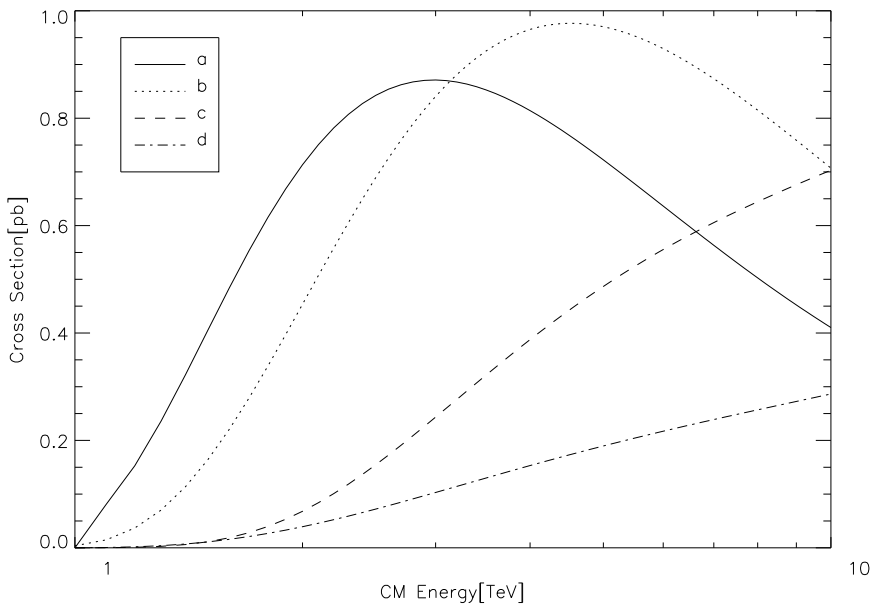}
  \caption{Total cross section versus CM energy of the incoming lepton(antilepton) and proton in the range of $1-10$ TeV.
  The lines a, b, c and d correspond to the reactions $l_{L}^{-}p\rightarrow \tilde{\nu}_{l}\tilde{d}X$, $l_{R}^{+}p\rightarrow \bar{\tilde{\nu}}_{l}\tilde{u} X$,
  $l_{L}^{-}p\rightarrow \tilde{\nu}_{l}\bar{\tilde{u}}X$ and $l_{R}^{+}p\rightarrow\bar{\tilde{\nu}}_{l}\bar{\tilde{d}}X$ respectively.}\label{Figure I}
\end{figure}

 For the signal we confine ourselves to the reaction
$ep\rightarrow \tilde\nu_ e\tilde d X $ only. According to the
Table I, decay modes of the scalar neutrino are
$\tilde\nu_e\rightarrow e^-\tilde\chi^+_1$,
$\tilde\nu_e\rightarrow \nu_e\tilde\chi^0_1$ and
$\tilde\nu_e\rightarrow e^-\tilde\chi^0_2$. Since R-parity is
conserved the lightest neutralino, $\tilde\chi^0_1$, is stable
whereas $\tilde\chi^0_1$ and $\tilde\chi^0_2$ have the following
leptonic and hadronic decays: $\tilde\chi^+_1\rightarrow
l^+\nu_e\tilde\chi^0_1$, $\tilde\chi^+_1\rightarrow q'\bar q
\tilde\chi^0_1$ and $\tilde\chi^0_2\rightarrow \bar f
f\tilde\chi^0_1, (f=l,q)$.\ Main decay modes of down squark are:
$\tilde d_L\rightarrow u\tilde\chi^-_1$ and $\tilde d_L\rightarrow
\tilde\chi^0_i d$ $(i=1,2,3,4)$. Subsequent decays of
$\tilde\chi^-_1$ are going to be the charge conjugate of
$\tilde\chi^+_1$ decays and $\tilde\chi^0_{3,4}\rightarrow
l^+\nu\tilde\chi^-_1$, $\tilde\chi^0_{3,4}\rightarrow
l^-\bar\nu\tilde\chi^+_1$, $\tilde\chi^0_{3,4}\rightarrow\bar q
q'\tilde\chi^{\pm}_1$, $\tilde\chi^0_3\rightarrow\bar f
f\tilde\chi^0_{1,2}$ and $\tilde\chi^0_4\rightarrow\bar f
f\tilde\chi^0_{1,2,3}$. Taking into account that
$Br(\tilde\nu_e\rightarrow e^-\tilde\chi^+_1)\approx 0.5$,
$Br(\tilde d_L\rightarrow u\tilde\chi^-_1)\approx 0.3$ and
$Br(\tilde\chi^+_1\rightarrow
e^+\nu_e\tilde\chi^0_1)=Br(\tilde\chi^+_1\rightarrow
\mu^+\nu_{\mu}\tilde\chi^0_1)\approx 1/9$ we see that the process
under consideration will give $\approx 50$ clear events of the
type $e^- l^+ l^- + jet + $$missing$ $energy$. Similar estimations
can be made for other processes in $ep$ and $\mu p$ collisions
shown in the Table II.

As a second example let us consider the production of
supersymmetric partner of right-handed neutrino. Recent
experimental results on neutrino oscillations force us to include
the right-handed neutrinos into the SM, so to include the
$\tilde{\nu}_{R}$ into the MSSM. In this case the "right"
sneutrino can be the lightest supersymmetric particle (LSP) [7].
In general, superpartners of left and right-handed neutrinos mix
to generate the mass eigenstates therefore one has to deal with a
$6\times6$ mass matrix for simplicity here we ignore the
interfamily mixings and we are left
\begin{center}
\begin{tabular}{c c}
$\tilde{\nu}_{1}^{l}=\cos\varphi_{l}\tilde{\nu}_{L}^{l}+\sin\varphi_{l}\tilde{\nu}_{R}^{l}$
\\
$\tilde{\nu}_{2}^{l}=-\sin\varphi_{l}\tilde{\nu}_{L}^{l}+\cos\varphi_{l}\tilde{\nu}_{R}^{l}$
\end{tabular}
\end{center}
for each family. Production cross sections for
$\tilde{\nu}_{1}^{l}$ and $\tilde{\nu}_{2}^{l}$ are obtained from
the multiplication of the values given in Table II and the factors
$\cos^{2}\varphi_{l}$ and $\sin^{2}\varphi_{l}$, respectively. If
the $\tilde{\nu}_{2}^{e}$ is the LSP, in the reaction
$ep\rightarrow\tilde{\nu}_{2}^{e} \tilde{d} X$ only $\tilde{d}$
will decay. For decay chain $\tilde{d}\rightarrow u
\tilde{\chi}_{1}^{-}\rightarrow u \mu^{-} \nu_{\mu}
\tilde{\chi}_{1}^{0}$ we expect roughly $230\sin^{2}\varphi_{e}$
events with topology $\mu_{-}+jet+$$missing$ $energy$ with
negligible SM background. The nonobservation of such an event will
correspond to the upper limit $\sim 0.05$ for
$\sin^{2}\varphi_{e}$.
\begin{table}
  \centering
  \caption{\label{Table I}  The SUSY spectrum for SUGRA Point 6 from ATLAS TDR (The
first and second generation squarks and sleptons are degenerate in
mass).}
\begin{tabular}{|c|c||c|c||c|c|}\hline
  Particles & Masses (GeV) & Particle & Masses (GeV) & Particle&
  Masses (GeV)
  \\\hline\hline
  $\tilde{g}$ & 540 & $\tilde{u}_{L}$ & 511 &  $\tilde{b}_{2}$ & 480
  \\\hline
  $\tilde{\chi}_{1}^{\pm}$ & 152 & $\tilde{u}_{R}$ & 498 & $\tilde{e}_{L}$ & 250 \\\hline

 $\tilde{\chi}_{2}^{\pm}$& 307 & $\tilde{d}_{L}$ & 517 &$\tilde{e}_{R}$ & 219
   \\\hline
  $\tilde{\chi}_{1}^{0}$ & 81 & $\tilde{d}_{R}$ & 498 &$\tilde{\nu}_{e}$ & 237
  \\\hline
  $\tilde{\chi}_{2}^{0}$ & 152 & $\tilde{t}_{1}$ & 365 &$\tilde{\tau}_{1}$ & 132
  \\\hline
  $\tilde{\chi}_{3}^{0}$ & 286& $\tilde{t}_{2}$ & 517 &$\tilde{\tau}_{2}$ & 259
  \\\hline
  $\tilde{\chi}_{4}^{0}$& 304 & $\tilde{b}_{1}$ & 390 & $\tilde{\nu}_{\tau}$ & 218 \\
  \hline
\end{tabular}
\vspace{3cm}
\end{table}
\begin{table}[h]
\centering
  \caption{\label{Table II} Total cross-section in pb for sneutrino
-squark production at Linac$\otimes$LHC}
$(\mathcal{L}=10^{33}cm^{-2}s^{-1})$ and $\mu$p
($\mathcal{L}=10^{32}cm^{-2}s^{-1}$) colliders.
\begin{tabular}{|c||c|c||c|c|}\hline
  Sub- & \multicolumn{2}{c||}{Linac$\otimes$LHC($\sqrt{S}=5.3$ TeV)} & \multicolumn{2}{c|}{$\mu$p collider ($\sqrt{S}=3$ TeV)}
  \\\cline{2-5}
   process& $\sigma(pb)$&$event/year$& $\sigma(pb)$ &$event/year$
   \\\hline\hline
  $l_{L}^{-}u_{L}\rightarrow \tilde{\nu}_{l}\tilde{d}$ & $0.70$ & $7000$ & $0.87$ &
  $870$ \\\hline
  $l_{R}^{+}d_{L}\rightarrow \bar{\tilde{\nu}}_{l}\tilde{u}$& $0.96$ & $9600$& $0.84$ &
  $840$ \\\hline
  $l_{L}^{-}\bar{d}_{R}\rightarrow \tilde{\nu}_{l}\bar{\tilde{u}}$ & $0.51$ & $5100$ & $0.24$ & $240$ \\ \hline
  $l_{R}^{+}\bar{u}_{R}\rightarrow\bar{\tilde{\nu}}_{l}\bar{\tilde{d}}$&$0.20$&$2000$&$0.10$&$100$\\\hline

\end{tabular}
\end{table}
\section{CONCLUSIONS}
Our estimations show that Linac$\otimes$LHC based ep collider has
great potential in searching for supersymmetric particles.
Although $\mu p$ collider has also considerable potential in
searching for SUSY, because of its less luminosity it is not
compatible with Linac$\otimes$LHC.

\end{document}